\documentclass[a4paper,oneside,12pt]{article}

\usepackage[T1]{fontenc}
\usepackage[latin1]{inputenc}
\usepackage[english]{babel}
\usepackage{lmodern} %Type1-font for non-english texts and characters
\usepackage{authblk}
\usepackage{url}

\usepackage{float}
\usepackage[dvips]{epsfig}

\usepackage{pdfpages}
\usepackage{graphicx, caption, subcaption}
\usepackage{epstopdf}
\usepackage[export]{adjustbox}

\usepackage[T1]{fontenc}
\usepackage[latin1]{inputenc}
\usepackage[english]{babel}

\usepackage{amsmath}
\usepackage{amsthm}
\usepackage{amsfonts}

\usepackage{ulem} %\sout{strike out}

\begin{document}

\title{Hyperchaos \& Labyrinth chaos: \\ Revisiting Thomas-R\"ossler systems}

\author[1]{Vasileios Basios}
\author[2]{Chris G. Antonopoulos}

\affil[1]{Service de Physique des Syst\`emes Complexes et M\'ecanique Statistique, Universit\'e 
Libre de Bruxelles, Interdisciplinary Center
for Nonlinear Phenomena and Complex Systems (CeNoLi), Brussels, CP231, Belgium}
\affil[2]{Department of Mathematical Sciences, University of Essex, Wivenhoe Park, CO4 3SQ, UK}

\date{\today}                     %% if you don't need date to appear
\setcounter{Maxaffil}{0}
\renewcommand\Affilfont{\itshape\small}

\maketitle

\begin{abstract}

We consider a multi-dimensional extension of the Thomas-R\"ossler (TR) systems, that was inspired by 
Thomas' earlier work on biological feedback circuits, and we report on our first results that shows 
its ability to sustain a spatio-temporal behaviour, reminiscent of chimera states. The novelty here 
being that its underlying mechanism is based on ``chaotic walks'' discovered by Ren\'e Thomas during 
the course of his investigations on what he called  Labyrinth Chaos. We briefly review the main 
properties of TR systems and their chaotic and hyperchaotic dynamics and discuss the simplest way of 
coupling, necessary for this spatio-temporal behaviour that allows the emergence of complex 
dynamical behaviours. We also recall Ren\'e Thomas' memorable influence and interaction with the 
authors as we dedicate this work to his memory. 

\end{abstract}

\pagestyle{plain} %Now display headings: headings / fancy / ...
\newpage

\section{Introduction}\label{sect:intro}
During the latter part of Ren\'e Thomas' brilliant scientific life, we had the opportunity to
collaborate with him researching on a class of models that himself and, his good friend and 
equally brilliant scientist, Professor Otto E. R\"ossler (of the ``R\"osler attractor'' fame) had 
proposed in the course of their investigations on the fundamentals of chaotic dynamics 
\cite{ThomRoss2004, ThomRoss2006}. 

Our interaction started by Ren\'e, ``naively'', asking questions 
and seeking assistance on tricks and tips for his favourite computational platform. He was always 
presenting the most profound and fundamental questions that he was working on, as if it was just 
his joyful occupation, now that he had retired, as he put it. You can imagine 
our enthusiasm when he asked if we can assist him and his friend Otto, with certain issues they were 
pondering at the time. From the knowledge we acquired from both, we now know, that it is one of those experiences most cherished. Besides Ren\'e's vast scientific 
knowledge and contributions, which this volume celebrates, our every-day interaction was a 
wonderful intellectual journey. Along with his scientific discourse of the highest quality 
(telling us about his circuits and loops, the necessary conditions of chaos; the 
utility of graph theory in studying dynamics much before the explosion of the work on 
network-dynamics; the importance of logic that underlies dynamics), Ren\'e would 
interweave the plethora of his intellectual passions. We learned from him about the contributions 
of amateur astronomers like himself; when things were gloomy he could always encourage us with 
stories and metaphors from his past climbing expeditions and activism. And of course, he would never cease 
to mention music even when he, so generously, shared his thoughts and work on some of his favourite 
toy-models, such as ``Labyrinth chaos'' and the ``Arabesques'' \cite{ThomRoss2004, Arabesque2013}.
By the way, we also learned  that Haydn, his favourite music composer, has a piece on chaos, called ``Die Vorstellung des Chaos''! Among all these, we naturally came to consider him as an 
epitome of the benefits of basic research. Always driven by his intellectual curiosity, Ren\'e opened 
roads that nobody else could see the possibility of them being opened; and he had such a great fun 
doing it! These are fond memories indeed, that will always stay with us to guide us in our future scientific endeavours.

Going back to Ren\'e's favourite toy-models: The purpose of this paper is to revisit and propose new 
directions that stem directly from his seminal investigations on hyperchaos \cite{ThomRoss2004, 
ThomRoss2006, Arabesque2013}.

Within the framework developed by him, M. Kaufman, D. Thieffry and coworkers 
\cite{Kaufman2013, Kaufman2001A, Kaufman2001B, Kaufman1995}, the dynamical basis of regulatory 
networks, cell differentiation, multistationarity, homeostasis and memory can be analysed and 
understood by studying theoretical models based on feedback circuits. The conceptual and analytic tools 
therein were also extended and can be used in the study of emergence of complex behaviour from simple circuit 
structures \cite{Kaufman2003complex}, not only in systems pertaining to biological models per se.
One of the seminal contributions of Ren\'e's work is his proposition of 
general rules in the dynamics of systems. In more details, (i) a positive circuit is necessary to 
display multiple stable states, and (ii) a negative circuit is necessary to have robust sustained 
oscillations. Latter on, Ren\'e proposed a necessary condition for chaos and 
suggested that both a positive and negative circuit are needed to generate deterministic 
chaotic behaviour. 

Moreover, Ren\'e and, O. E. R\"ossler and coworkers \cite{ThomRoss2004}, 
have shown further that for dimensions $D\geq 4$, the same simple logical 
structure of more than one positive and at least one negative circuit can generate ``hyperchaos''
of arbitrary order $m$, i.e., chaotic behaviour characterised by more than $m$ positive Lyapunov exponents. 
As well known, among Ren\'e's contributions to the theory of dynamical systems (for details see also references within this
volume), are ``feedback circuits'' or simply ``circuits'', defined as sets of nonzero terms of 
the Jacobian matrix (linearisation) of the dynamical system such that their row  and column indices can form cyclic permutations of each other. In that sense, hyperchaos of order $m$ requires the existence of $m$ positive circuits and at 
least one negative. As elegant as this!

Ren\'e and O. E. R\"ossler considered the generation of complex symmetric attractors (which they termed ``Labyrinth chaos")\cite{ThomRoss2004, ThomRoss2006} and its peculiar special case of a chaotic phase-space where a countable-infinite set of unstable fixed points with no attractors is generated by a set of $m>=3$ first 
order differential equations. Based on such systems, Ren\'e then created a class of conservative systems which he termed ``Arabesques'' \cite{Arabesque2013}.

The discovery and, subsequent definition of hyperchaos \cite{Hyperchaos79, Scholarpedia} is due to the seminal 
work of O. E. R\"ossler, and describes a more flexible type of chaotic behaviour where the 
sensitive dependence on initial conditions co-exists in more than one directions. In other words, there are more than one positive Lyapunov exponents in the dynamics of such systems. Hyperchaos has been studied initially in experimental works on systems with coupled lasers \cite{Hyperchaos88}, to Navier-Stokes equations \cite{HyperchaosNavier}, to recent studies of large arrays of coupled oscillators \cite{Hyperchaos17}. Surprisingly, the higher dimensional chaotic motion is more amenable to control and synchronisation 
and quite evidently empowers the design of controllers in circuit theory and 
applications \cite{HyperchaosDesign}. It also has good utility in studies in biological 
modelling in areas such as Bioinformatics \cite{HyperchaosGenes}, biorythms and chronotherapy \cite{HyperchaosChrono}, and neuronal dynamics \cite{HyperchaosNeurons}.

In Section \ref{sect:2}, we revisit the system of ordinary differential equations for hyperchaos
introduced initially by Ren\'e and O. E. R\"ossler \cite{ThomRoss2004, ThomRoss2006}. Since they called this type of 
dynamics ``Labyrinth chaos'', we shall use the term Thomas-R\"ossler (TR) systems, as it has been proposed in the literature \cite{LiSprott2014} to refer to them, and review its basic properties. In Section \ref{sect:3}, we extent the original investigations by considering first linearly coupled systems of two and then, of a larger number of RT systems arranged in networks in a circle with nearest-neighbour interactions. In Section \ref{sect:conclusion}, we conclude and discuss briefly possible future research based on these investigations and provide an outlook of the fruitful continuation of this particular direction of research that was also initiated by Ren\'e Thomas \cite{ThomRoss2004, Arabesque2013}.

\section{Labyrinth Chaos \& Hyperchaos} \label{sect:2}

In \cite{ThomRoss2004}, Ren\'e, O. E. R\"ossler and co-workers proposed a system of coupled ordinary 
differential equations to elucidate, in terms of feedback circuits, the necessary conditions for chaotic 
and hyperchaotic motion. They considered the following equations
\begin{equation}\label{eq:ThomRoss}
 \frac{dx_i}{dt} = -b x_i + f(x_{i+1})  ,\qquad  i=1\ldots N\;\;(\mbox{mod } N),
\end{equation}
where $0<b<1$.
The circuits of this system can be positive or negative depending on the location in the phase space. This constitutes an ``ambiguous'' circuit, in the terminology established by Ren\'e \cite{Kaufman1995, Kaufman2001A, Kaufman2001B}. As it had already been shown, under proper conditions, a single circuit might be 
sufficient to generate chaotic dynamics. Ren\'e and O. E. R\"ossler confirmed and generalised this proposition, and showed 
that hyperchaos of order $m>1$ can be generated by a single ambiguous circuit of 
dimension $2m$. The function $f$ was taken to be nonlinear. In this case, either 
$f(u)=u^3-u$ or $f(u)=\sin(u)$ and used $N=3$ for chaos, and $N=5$ for hyperchaos. 
Ren\'e termed a special class of such systems, with $b=0$ and $f$ assumming different 
forms, as ``Arabesques'' (for a detailed discussion, see \cite{Arabesque2013}).   

Let us focus now on the case where $f(u)=\sin(u)$ and $N=3$. The 3-dimensional version of the system then reads
%%%%%%%%%%%%%%%%%%%%%%%%%%%%%%%%%%%%%%%%%%%%
\begin{eqnarray}\label{eq:3D-ThomRoss}
\frac{dx}{dt} &=& -b x  +\sin(y)\nonumber\\
\frac{dy}{dt} &=& -b y  + \sin(z)\\
\frac{dz}{dt} &=& -b z  + \sin(x),\nonumber
\end{eqnarray}
and its jacobian is given by
\begin{equation}\label{eq:3D-Jacobian}
\bold{J}=\left[ \begin {array}{ccc} -b&\cos \left( y_{{1}} \right) &0
\\\noalign{\medskip}0&-b&\cos \left( z_{{1}} \right) 
\\\noalign{\medskip}\cos \left( x_{{1}} \right) &0&-b\end {array}
 \right].
\end{equation}
%%%%%%%%%%%%%%%%%%%%%%%%%%%%%%%%%%%%%%%%%%%%%

\begin{figure}[H]
  \centering
 % \graphicspath{./figs}
 %\includegraphics[scale=0.15]{b-vs-LEs-3Dsin.png}(a)
 \begin{subfigure}{0.48\textwidth} % width of left subfigure
 \centering   \includegraphics[width=0.97\textwidth]{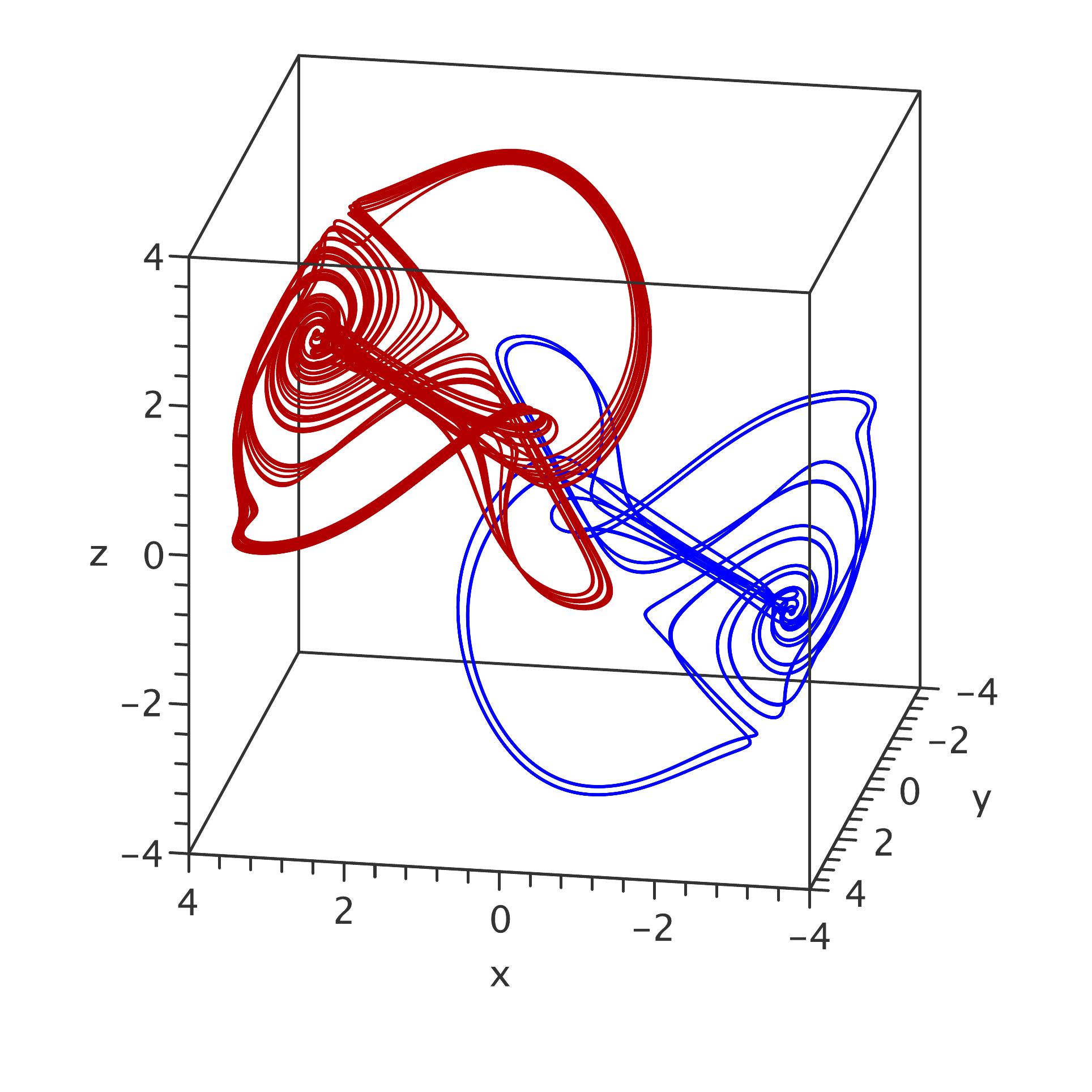}
  \end{subfigure}
 \begin{subfigure}{0.48\textwidth} % width of left subfigure
 \centering     \includegraphics[width=0.97\textwidth]{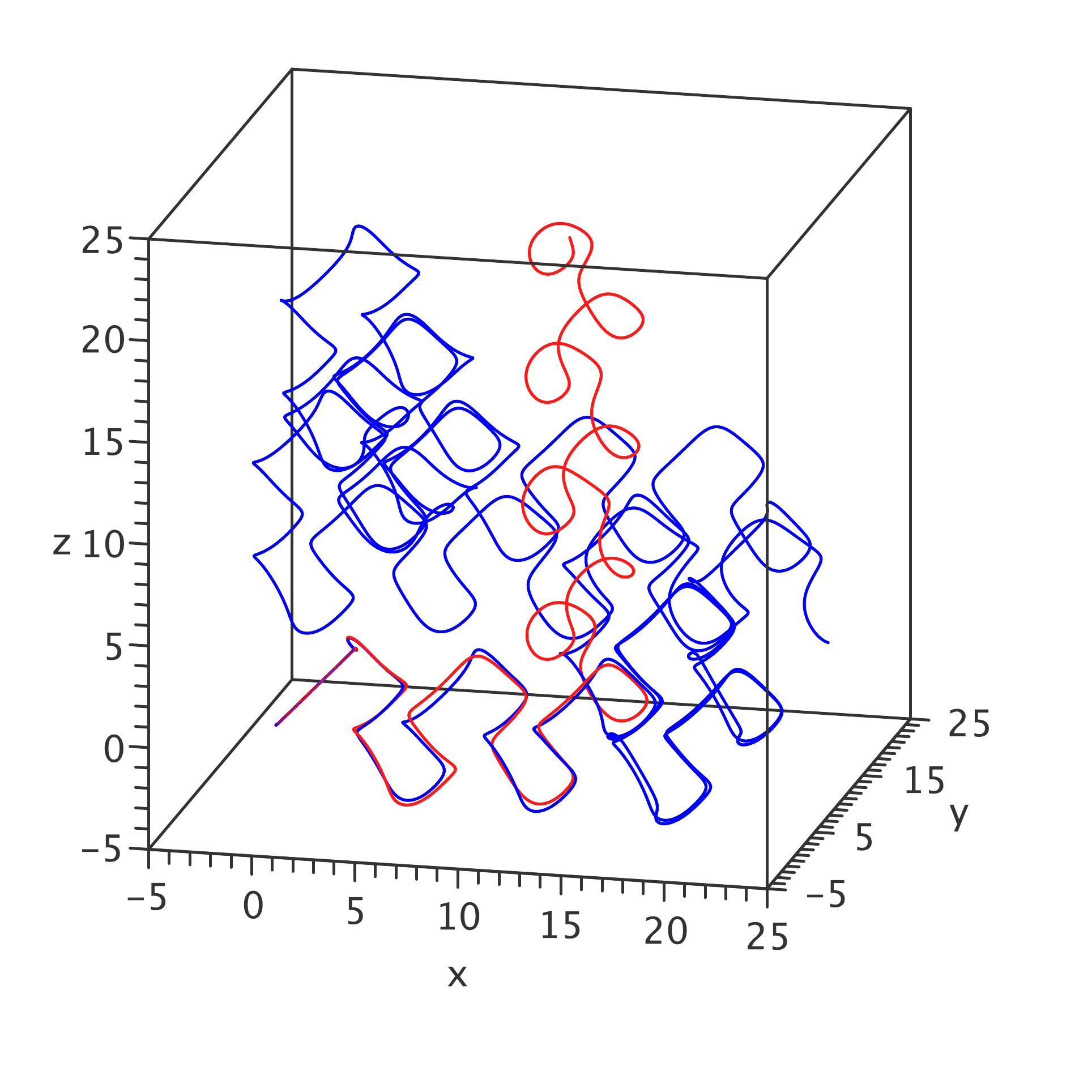}
  %\llap{\raisebox{5.0cm}{\includegraphics[height=3.0cm]{b-vs-LEs-3Dsin.png}}}
  \llap{\raisebox{5.0cm}{\includegraphics[height=3.0cm,frame]{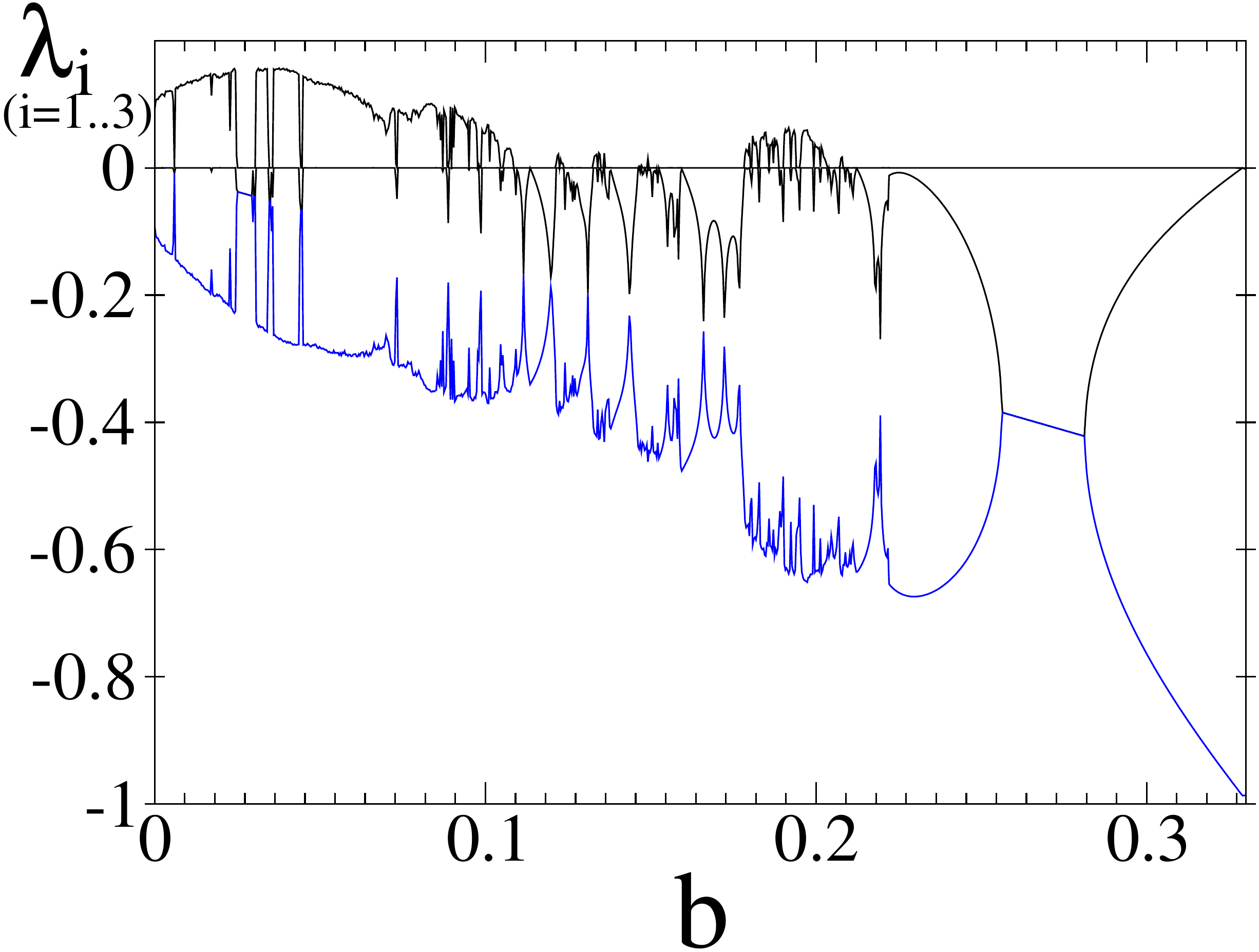}}}
         \caption{}
  \end{subfigure}
  \caption{Chaotic attractors and labyrinth chaos in the 3-dimensional version of the system of Eq. \eqref{eq:ThomRoss}. (a) Two coexisting attractors for $b=0.2$. (b) ``Chaotic walks and labyrinth chaos'' for $b=0$ for which the system is conservative. The red and blue trajectories exhibit sensitive dependence on initial conditions and diverge in time. In the inlet, the dependence of the 3 Lyapunov exponents as a function of $0\leq b <1/3$ (see \cite{ThomRoss2004}).}
  \label{fig:sin3D}
\end{figure}

System \eqref{eq:3D-ThomRoss} exhibits a rich repertoire 
of dynamical behaviours for different $b>0$ values \cite{ThomRoss2004}. For example, for $b=0.19$ we have a complicated but stable periodic 
orbit, and for $b=0.18$, a single chaotic attractor. Also, more complex dynamical regimes can appear. For 
example, for $b=0.2$, the system possesses two co-existing chaotic attractors, which we show in 
Fig. \ref{fig:sin3D}(a). In the special case where $b=0$, quite an
exotic chaotic behaviour appears, without any attractors. In this case, system \eqref{eq:3D-ThomRoss} is conservative with an infinite lattice of unstable fixed points. The behaviour shown in Fig. \ref{fig:sin3D}(b) is what Ren\'e termed ``chaotic walks in labyrinth chaos''.

%%%%%%%%%%%%%%%%%%%%%%%%%%%%%%%%%%%%%%%%%%%%
%%%%%%%%%%%%%%%%%%%%%%%%%%%%%%%%%%%%%%%%%%%%%%%%%%%%%%%%%%%%%%%%%%

Similar behaviour appears for $N=5$ \cite{ThomRoss2004} as well. A very detailed study of this system 
was undertaken in \cite{SprottChlouverakis} and it is presented as a prototypical example of 
chaos in \cite{SprottBook}. As it is noted in  \cite{SprottChlouverakis}, ``Despite its 
mathematical simplicity, this system of ordinary differential equations produces
a surprisingly rich dynamic behaviour that can serve as a prototype for chaos studies''. They also remark 
that in the case of chaotic walks, the approach of an ensemble of initial 
conditions to equilibrium is by way of fractional Brownian motion with a Hurst exponent of approximately $0.61$ and a slightly leptokurtic distribution. To the best of our knowledge, this might be the only example of a simple system that links fractional-Brownian motion with nonlinear feedback. 

Finally, let us note that the simple, underlying structure of the 
TR system provides for the equally simple and elegant form of the characteristic equation of its Jacobian matrix $\bold{J}$. In this case, the $N\times N$ Jacobian of Eq. \eqref{eq:ThomRoss}, for a sinusoidal nonlinearity (i.e. for $f(u)=\sin(u)$), reads
\begin{equation}\label{eq:NDcharacteristic}
\det{(\bold{J}-e_i \bold{I})} = (-1)^N\left[{(b-e_i)}^{N}-\prod_{i=1}^N\cos \left( x_{{i}} \right) 
\cos \left( y_{{i}} \right) \cos \left( z_{{i}} \right)  \right],
\end{equation}
where $e_i,\;i=1,\ldots,N$ are the $N$ eigenvalues of $\bold{J}$. Since only those terms in $\bold{J}$ that belong to one or more circuits are represented in the characteristic equation and hence take part in the calculation of 
the eigenvalues, $e_i$, it bears significant effect on the calculation of the Lyapunov exponents $\lambda_i$ of the system.

\section{Revisiting the Thomas-R\"ossler class of systems}\label{sect:3}
\subsection{Two linearly coupled 3-dimensional Thomas-R\"ossler systems} 
Motivated by the above, we turn here to the question of the effect the simplest {\it linear} 
coupling has to two 3-dimensional TR systems, ($N=6$), and how that 
compares to the hyperchaotic case of equal dimensions $N=2\times3=6$. As we shall see below, the linearly coupled $6$-dimensional system can produce hyperchaotic behaviour. Yet, this is due to a different underlying logic of its feedback circuits.

Obviously, the simplest way to couple two such 3-dimensional systems is by a linear coupling involving their $x$ variables. This is considering two copies of Eq. \eqref{eq:3D-ThomRoss}
\begin{eqnarray}\label{eq:2X3D-ThomRoss}
\frac{dx_{1,2}}{dt} &=& -b_{1,2} x_{1,2}  +\sin(y_{1,2}) + \frac{d}{2} (x_{1,2} - x_{2,1})\nonumber\\
\frac{dy_{1,2}}{dt} &=& -b_{1,2} y_{1,2}  +\sin(z_{1,2}),\\
\frac{dz_{1,2}}{dt} &=& -b_{1,2} z_{1,2}  +\sin(x_{1,2})\nonumber
\end{eqnarray}
where $d\geq0$ and $b_{1,2}\geq0$.
The coupling has a direct effect on the structure of the Jacobian $\bold{J}$ of the coupled system. If $\bold{J}_1$ and $\bold{J}_2$ are the Jacobian matrices of the two 3-dimensional copies, then the Jacobian $\bold{J}_d$ of the coupled system of Eq. \eqref{eq:2X3D-ThomRoss} is the $6\times6$ matrix
\begin{equation}\label{eq:2X3Dcharacteristic}
 \bold{J}_d=\left[ 
 \begin {array}{cc} 
 \bold{J}_1 & \bold{D} \\
 \bold{D} & \bold{J}_2
\end {array}
 \right]
 \mbox{, where }
\bold{D}=\left[ \begin {array}{ccc} -\frac{1}{2}d&0&0\\\noalign{\medskip}0&0&0
\\\noalign{\medskip}0&0&0\end {array} \right].
\end{equation}\label{eq:2X3D}
The characteristic equation of $\bold{J}_d$ then reads
\begin{equation}
 \det{(\bold{J}_c-e_i \bold{I})} = \det(\bold{J}_2-e_i \bold{I})\cdot
\det((\bold{J}_1-e_i \bold{I})-\bold{D} \bold{J}_2^{-1} \bold{D})
\end{equation}
using the Shur complement for the blocks of matrix $\bold{J}_c$, where $\bold{I}$ is the $6\times6$ identity matrix. Evidently,
Eq. \eqref{eq:2X3Dcharacteristic} cannot be reduced via Eq. \eqref{eq:NDcharacteristic} for any value 
of $N$. Even setting $b_1=b_2=b$, the simplification is insignificant as in  
Eq. \eqref{eq:2X3Dcharacteristic} we encounter all terms of $b_{1,2},e_i,d$, and all their powers
combined up to order $N$, as well as the products of the cosines of all variables in 
an irreducible manner.

%%%%%%%%%%%%%%%%%%%%%%%%%%%%%%%%%%%%%%%%%%%%%%%%
\begin{figure}[H]
  \centering
% \graphicspath{./figs}
\begin{subfigure}{0.48\textwidth} % width of left subfigure
 \centering
 \includegraphics[width=1.3\textwidth]{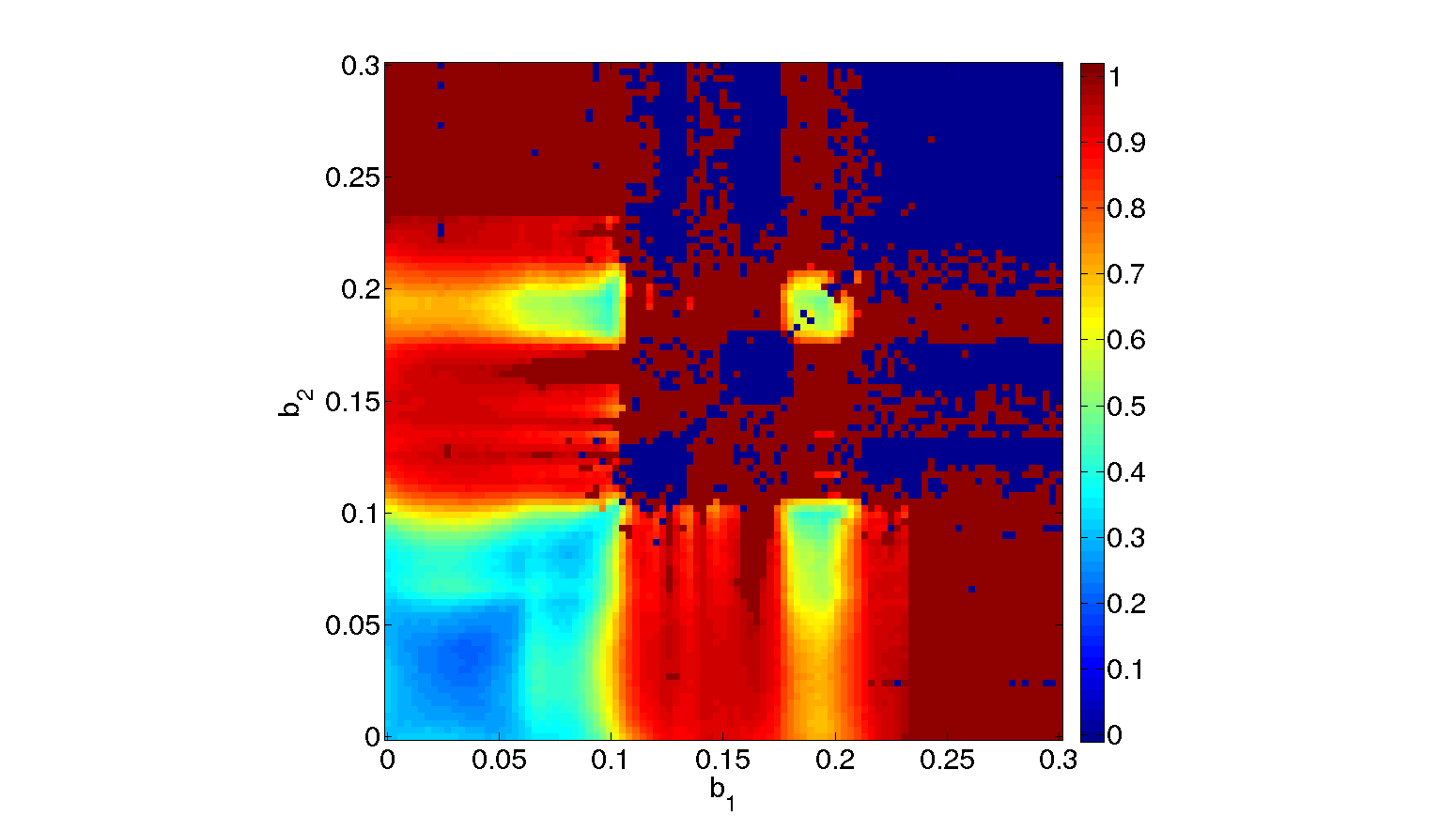}
         \caption{}
  \end{subfigure}
  \begin{subfigure}{0.48\textwidth} % width of left subfigure
 \centering \includegraphics[width=1.3\textwidth]{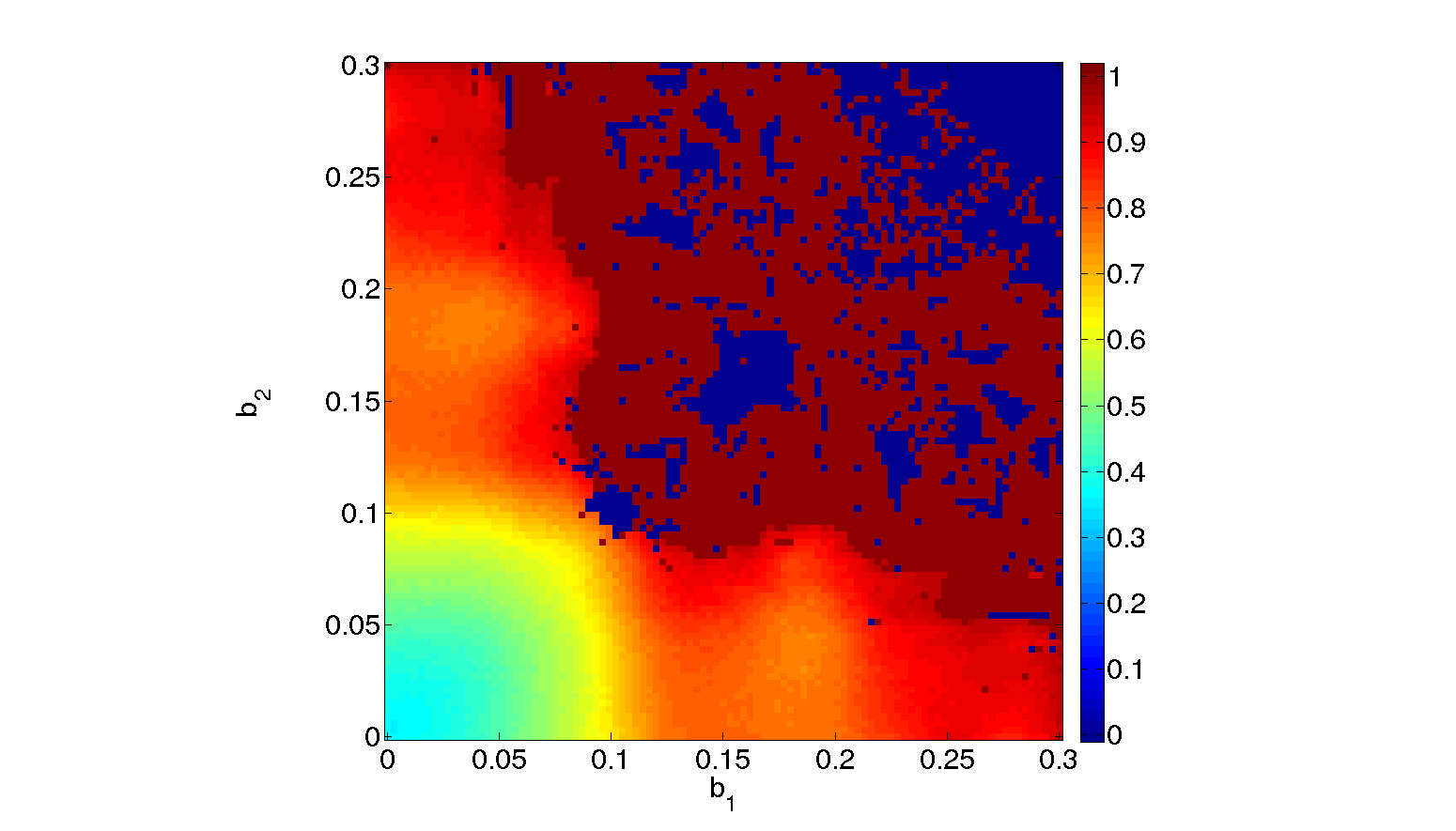}
         \caption{}
  \end{subfigure}
  \begin{subfigure}{0.48\textwidth} % width of left subfigure
 \centering \includegraphics[width=1.3\textwidth]{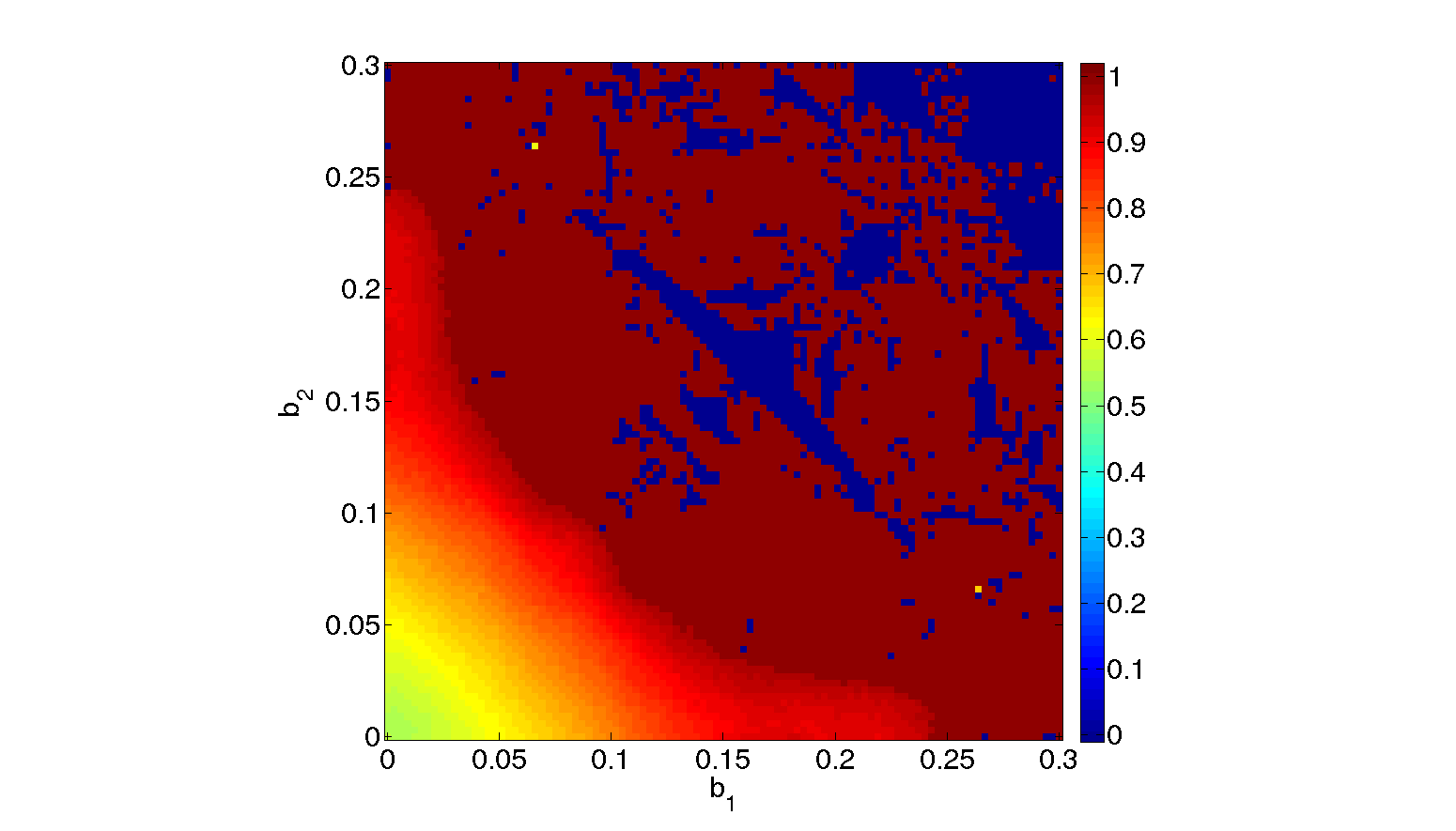}
         \caption{}
  \end{subfigure}
  \caption{The parameter space $(b_1, b_2)$ for two linearly coupled 3-dimensional TR systems of Eq. \eqref{eq:2X3D-ThomRoss}. The colour-code denotes the relative difference $\Delta\lambda$ (see text) between the first two positive Lyapunov exponents $\lambda_{1}$ and $\lambda_2$. Where these Lyapunov exponents are negative, they are assigned the value zero (depicted by dark blue). Hyperchaotic regions are depicted by the colours that correspond to $\Delta\lambda$ in between 0 and 1. Light blue: weaker hyperchaos and green to red: stronger hyperchaos. The coupling values were chosen as following: (a) $d=0.01$; (b) $d=0.1$; (c) $d=0.3$.} \label{fig:b1b2}
\end{figure}
%%%%%%%%%%%%%%%%%%%%%%%%%%%%%%%%%%%%%%%%%%%%%%

As is well known, the computation of the Lyapunov exponents $\lambda_i$ of a flow is based on the 
evaluation of an infinite, or sufficiently large for numerical purposes, product of their Jacobian 
matrices along their trajectories \cite{LiSprott2014} or equivalently, for a sufficiently long integration time \cite{LEs1980}. Figure \ref{fig:b1b2} shows a measure of the relative difference $\Delta\lambda = \frac{\lambda_{1}-\lambda_2}{\lambda_{1}}$ of the first two largest Lyapunov exponents $\lambda_{1}$ and $\lambda_{2}$ of Eq. \eqref{eq:2X3D-ThomRoss}. One can see that as $d$ increases (from panel (a) to (c)), the dark blue region where the Lyapunov exponents are not-positive, recedes. This means, the larger the coupling strength, the larger the hyperchaotic region in the parameter space $(b_1,b_2)$ (see Fig. \eqref{fig:b1b2} for more details).

\subsection{Can $N$ linearly coupled 3-dimensional Thomas-R\"ossler systems support chimera-like states?}

A remarkable novel discovery in the area of nonlinear dynamics and chaos was made by Kuramoto and Battogtokh in 2002 \cite{kuramotoetal2002} when they discovered the coexistence of coherent and incoherent behaviour in populations of non-locally coupled phase oscillators. As remarkable as counter intuitive it might be, their seminal work has triggered since then a fascinating interest for an ever growing research community witnessed by an almost exponential growth of publications, spanning the areas of physics, biology and mathematics, among others. As such states later came to be named ``chimera states'' \cite{chimera_Abrams}, this new phenomenon of synchronisation still lacks a complete and rigorous mathematical definition. Chimera state can be defined as spatio-temporal patterns in networks of coupled oscillators in which synchronous and asynchronous oscillations coexist. This state of broken symmetry, which usually coexists with a stable spatially symmetric state, has intrigued the nonlinear dynamics community since its discovery in the early 2000s \cite{kuramotoetal2002}. Nevertheless, recent experiments and its relevance to biological networks keeps an unceasing interest in the origin and dynamics of such states (see \cite{BountisIranians2018, chimera-neurons} and references therein). For a recent review on chimera states we refer the reader to \cite{Panaggioetal2015}.

\begin{figure}[ht!]
\centering
 \begin{subfigure}{0.49\textwidth} % width of left subfigure
 \centering   \includegraphics[width=0.8\textwidth,angle=-90]{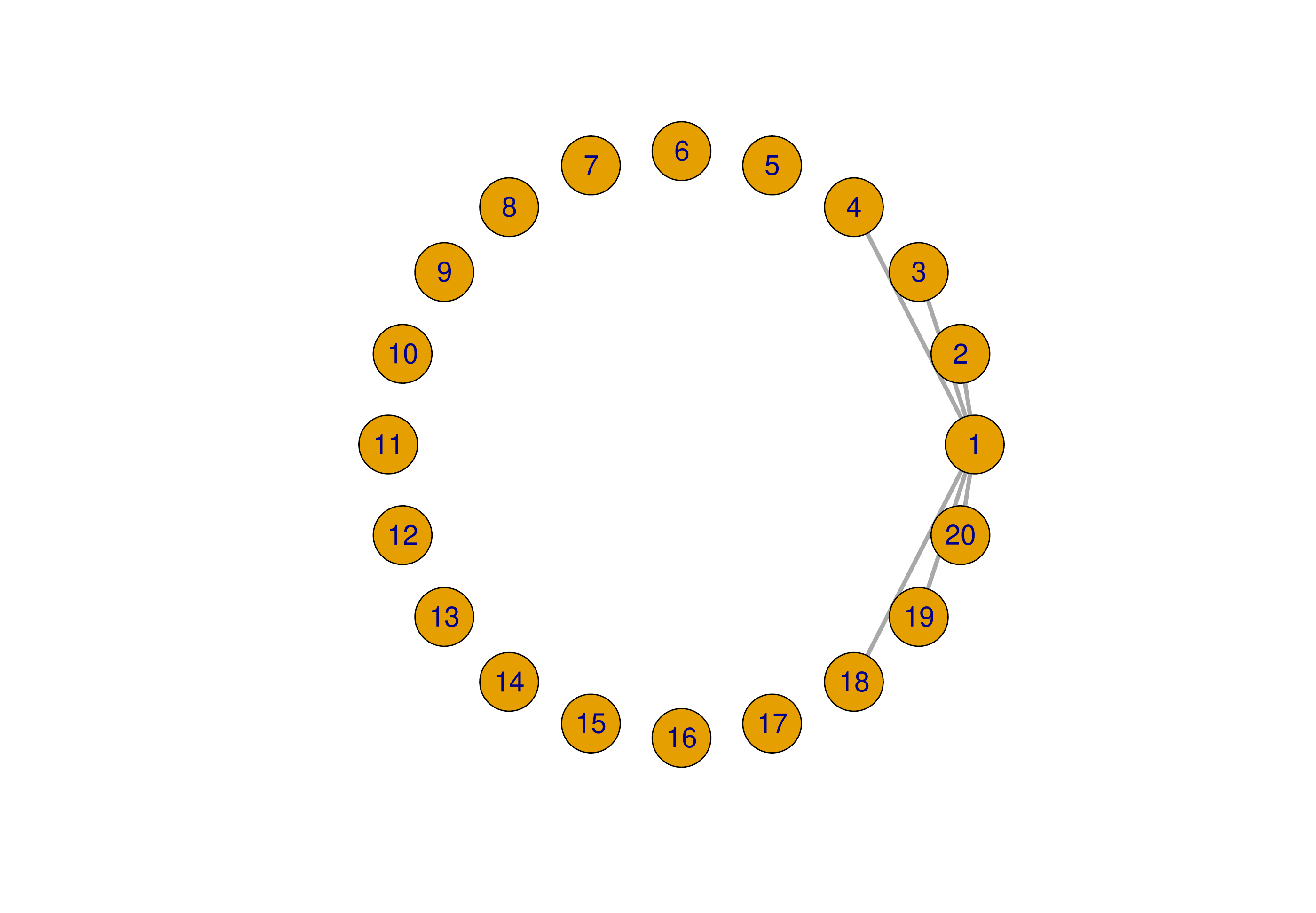}
         \caption{}
  \end{subfigure}
 \begin{subfigure}{0.49\textwidth} % width of left subfigure
 \centering   \includegraphics[width=0.8\textwidth,angle=-90]{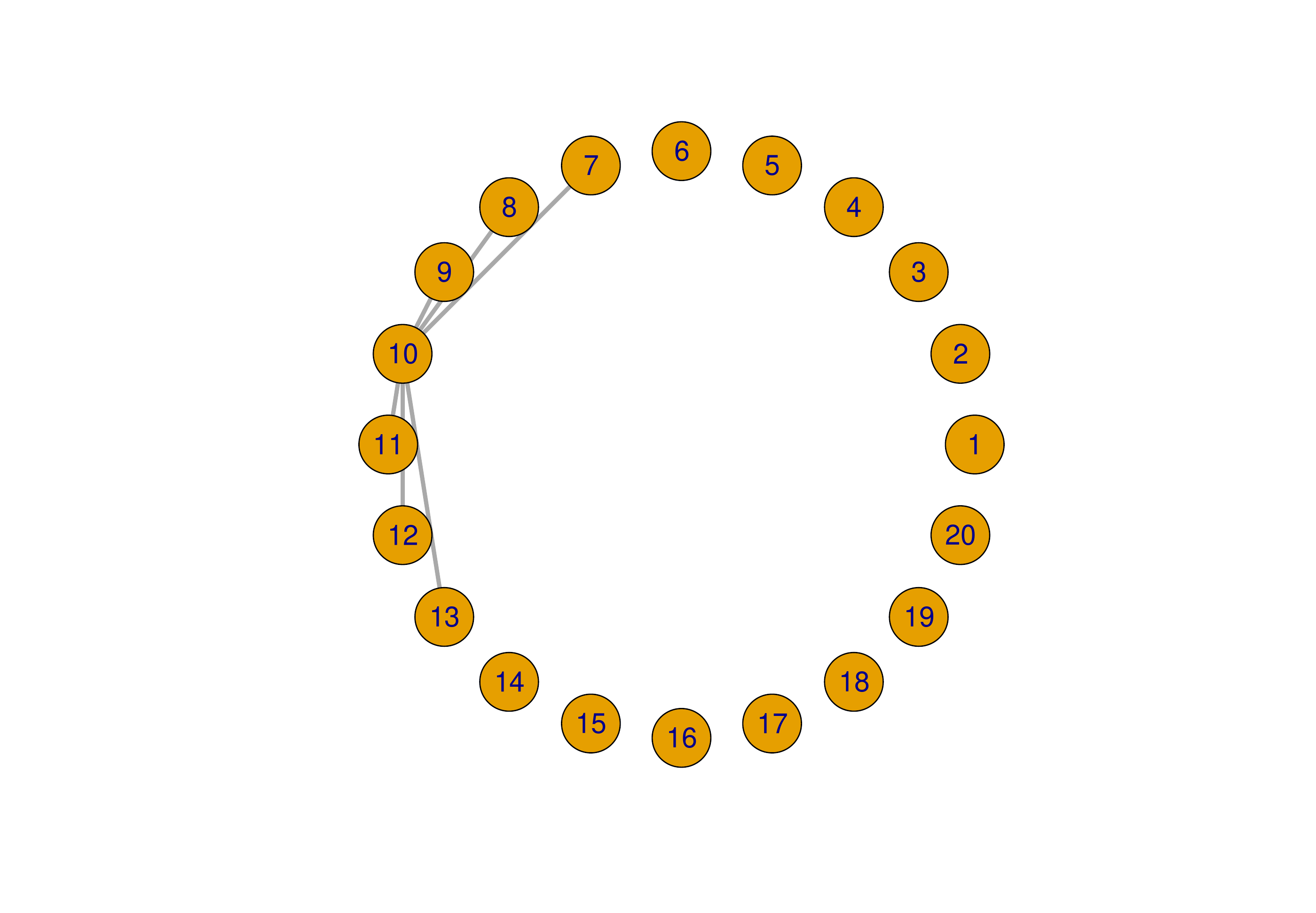}
         \caption{}
  \end{subfigure}
 \begin{subfigure}{0.49\textwidth} % width of left subfigure
 \centering   \includegraphics[width=0.8\textwidth,angle=-90]{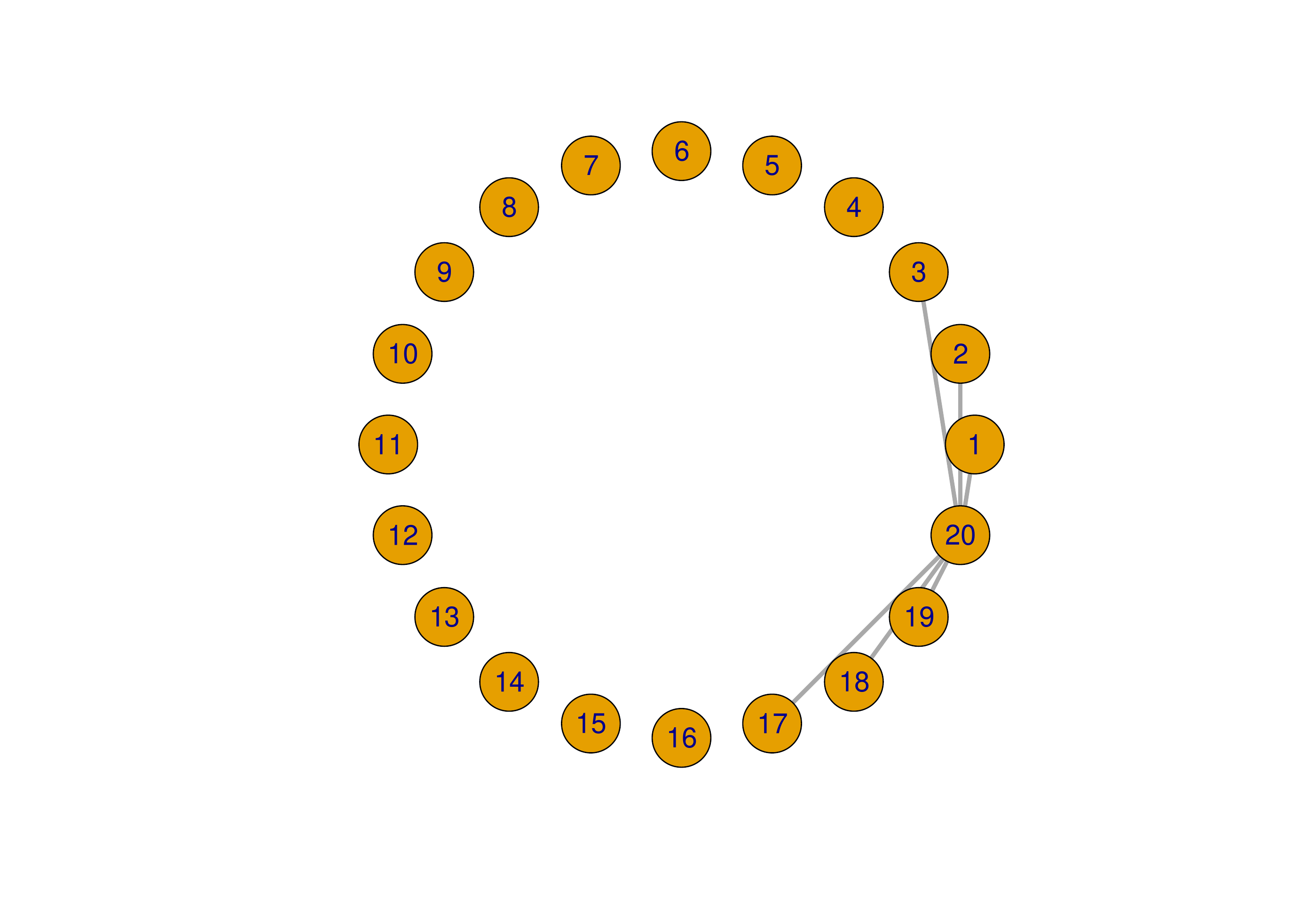}
         \caption{}
  \end{subfigure}
\caption{Three examples of a network topology with $N=20$ nodes arranged in a circle where each node is connected with $P=3$ nearest-neighbours, 3 in either side of the node. For illustration purposes, we only show the 6 nearest neighbours of node 1 in (a), of node 10 in (b) and of node 20 in (c), and is similar for all other nodes in the network, namely all nodes are connected with their 6 nearest neighbours. Notice the periodic boundary conditions for the first (panel (a)) and 20th nodes (panel (c)) in the network.}
\label{fig:Topology}
\end{figure}

Chimera states as phenomena of spatio-temporal patterns in networks of coupled oscillators have three generic characteristics: (i) nonlocal coupling, (ii) robust but varying coherent-incoherent patterns in both space and time (spatio-temporal patterns in which phase-locked oscillators coexist with drifting ones) and (iii) broken symmetry coexisting with a stable spatially symmetric state that depends on initial conditions as well as on system's parameters. So far, perfect or imperfect \cite{chimera-Imperfect} chimera states have not been detected for  just local or global coupling but are typical of the intermediate case: a nonlocal coupling comprising of a significant number of nearest neighbours, similar to the situation depicted in Fig. \ref{fig:Topology}.

Here, we present some preliminary results for $N$ linearly coupled TR systems (see Eq. \eqref{eq:3D-ThomRoss}) which are the generalisation of the system in Eq. \eqref{eq:2X3D-ThomRoss}, that provide evidence they can exhibit spatio-temporal phenomena of coherent and incoherent patterns that alternate dynamically in time, reminiscent of chimera states in networks of non-locally coupled oscillators \cite{kuramotoetal2002,chimera_Abrams}.

In particular, the system of $N$ linearly coupled TR systems considered is given by
%%%%%%%%%%%%%%%%%%%%%%%%%%%%%%%%%%%%%%%%%%%%%%%%%

\begin{eqnarray}\label{eq:3DNPThomRoss}
\frac{d x_k}{dt} &=& -b_k x_k +\sin(y_k)+\frac{d}{2P}\sum_{j=k-P}^{j+P}(x_k - x_j)\nonumber\\
\frac{d y_k}{dt} &=& -b_k y_k + \sin(z_k),\\
\frac{d z_k}{dt} &=& -b_k z_k + \sin(x_k)\nonumber
\end{eqnarray}
where $k=1\dots N$, $2P\leq N$, $d\geq 0$ is the strength of the linear coupling and $b_k\geq0$ for all $k$. The term $1/(2P)$ is a normalisation constant for the linear coupling among the $N$ 3-dimensional TR systems and the connectivity can be visualised by a similar network as in Fig. \ref{fig:Topology}.

Depending on the $b_k$ values, non-locality properties (number of nearest neighbours in either side of each node $2P$ should be close to $N$) and appropriate initial conditions, the system can exhibit fascinating, spatio-temporal phenomena of coherent and incoherent behaviour that alternate dynamically in time and are reminiscent of chimera states observed in networks of non-locally coupled oscillators \cite{kuramotoetal2002,chimera_Abrams,Panaggioetal2015}.

In the next, we study two interesting cases where: (a) the first half of the TR systems are conservative, exhibiting labyrinth chaos (i.e. $b_k=0$ for $k=1,\ldots,N/2$) and hyperchaos (i.e. $b_k=0.18$ for $k=(N/2)+1,\ldots,N$) when uncoupled and (b) the first half are conservative again and the rest half exhibit complex periodic oscillations (i.e. $b_k=0.19$ for $k=(N/2)+1,\ldots,N$) when uncoupled. These specific values of $b_k$ where taken from \cite{ThomRoss2004}.

\begin{figure}[ht!]
\centering
 \begin{subfigure}{0.49\textwidth} % width of left subfigure
 \centering   \includegraphics[width=0.8\textwidth]{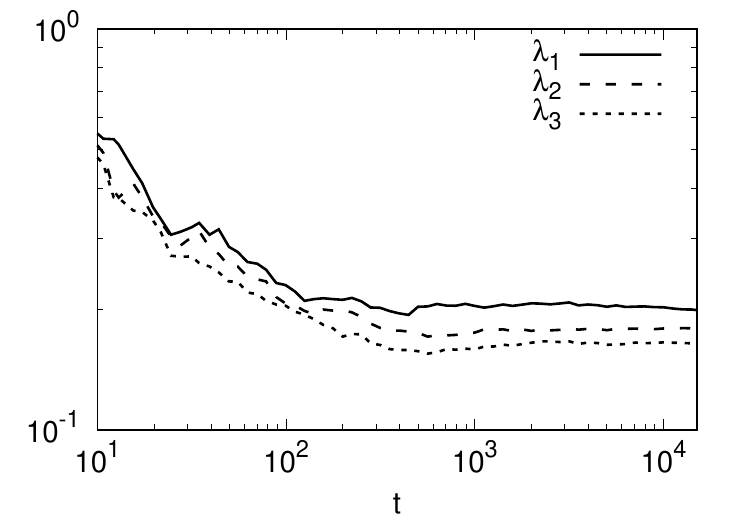}
         \caption{}
  \end{subfigure}
 \begin{subfigure}{0.49\textwidth} % width of left subfigure
 \centering   \includegraphics[width=0.8\textwidth]{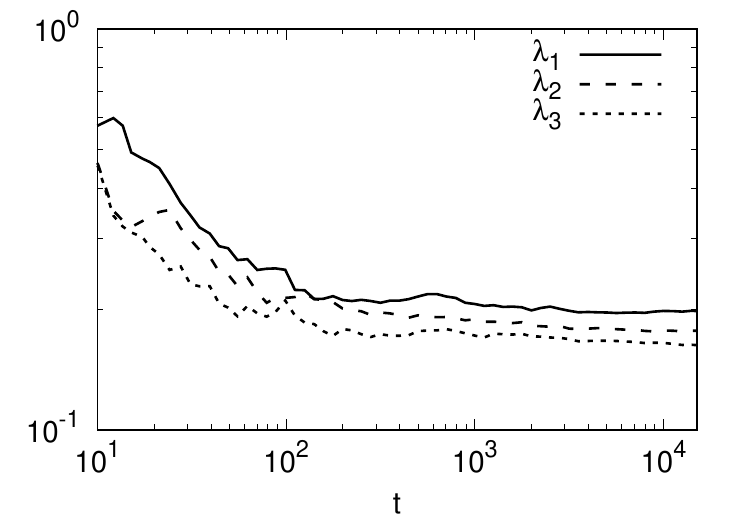}
         \caption{}
  \end{subfigure}
\caption{Hyperchaotic behaviour in a system of 40 linearly coupled 3-dimensional TR systems. The time-evolution of the first 3 largest Lyapunov exponents $\lambda_1$, $\lambda_2$ and $\lambda_3$ for a network of $N=40$ 3-dimensional linearly coupled TR systems as in Eq. \eqref{eq:3DNPThomRoss}, where each node is connected to its $P=20$ nearest neighbours, in either side of each node. The system resides in a hyperchaotic regime in both cases as more than one Lyapunov exponents are non-zero and  positive. In (a) $b_k=0$ for $k=1,\ldots,N/2$) and $b_k=0.18$ for $k=(N/2)+1,\ldots,N$ (labyrinth chaos and hyperchaos) and in (b) $b_k=0$ for $k=1,\ldots,N/2$) and $b_k=0.19$ for $k=(N/2)+1,\ldots,N$ (labyrinth chaos and complex periodic oscillations). The final integration time is $t=1.5\times10^4$ where convergence of the Lyapunov exponents to positive, non-zero, values is evident.}
\label{fig:LEs_Network}
\end{figure}

In both cases, when these systems are coupled together and run for sufficiently long integration times, they exhibit hyperchaotic behaviour as manifested by the convergence of more than 1 Lyapunov exponents to positive, non-zero, values in the course of time. We show this in fig. \ref{fig:LEs_Network}, where we plot the first 3 Lyapunov exponents as a function of time (final integration time is $1.5\times10^4$) for a system of $N=40$ TR 3-dimensional systems with $P=20$ (to guarantee non-locality in the connectivity). In both cases, these Lyapunov exponents show a clear tendency to converge to positive, non-zero values.

%%%%%%%%%%%%%%%%%%%%%%%%%%%%%%%%%%%%%%%%%%%%%%%%%%
  
\subsection{Chimera states in a multi-dimensional TR system with labyrinth chaos and hyperchaos}

\begin{figure}[ht!]
\begin{subfigure}{0.4\textwidth}
\includegraphics[width=1\textwidth]{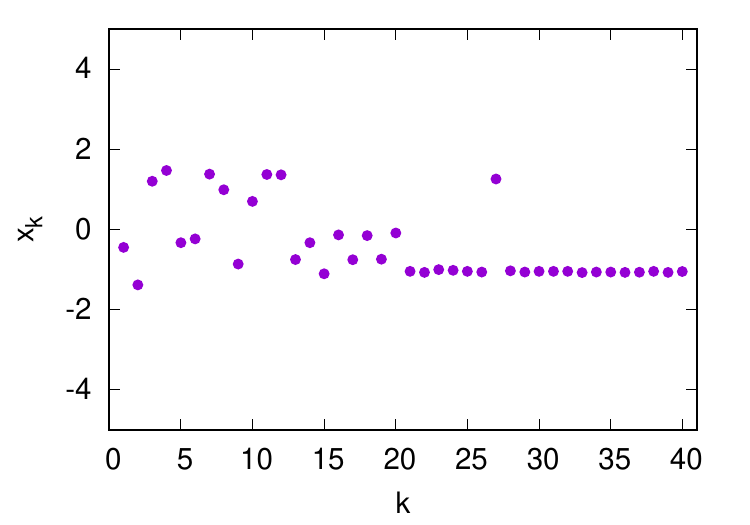}\\
\includegraphics[width=1\textwidth]{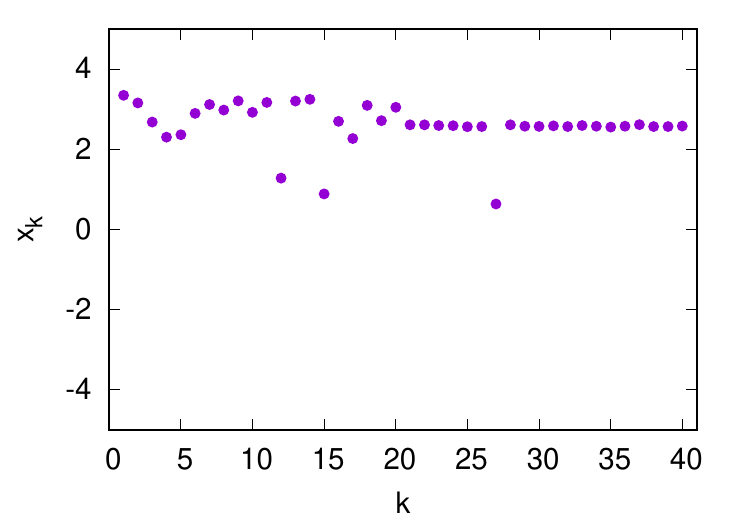}
\caption{}
\label{fig:profile_b00b018}
\end{subfigure}
\begin{subfigure}{0.5\textwidth}
\includegraphics[width=9.7cm,height=10cm,angle=-90]
{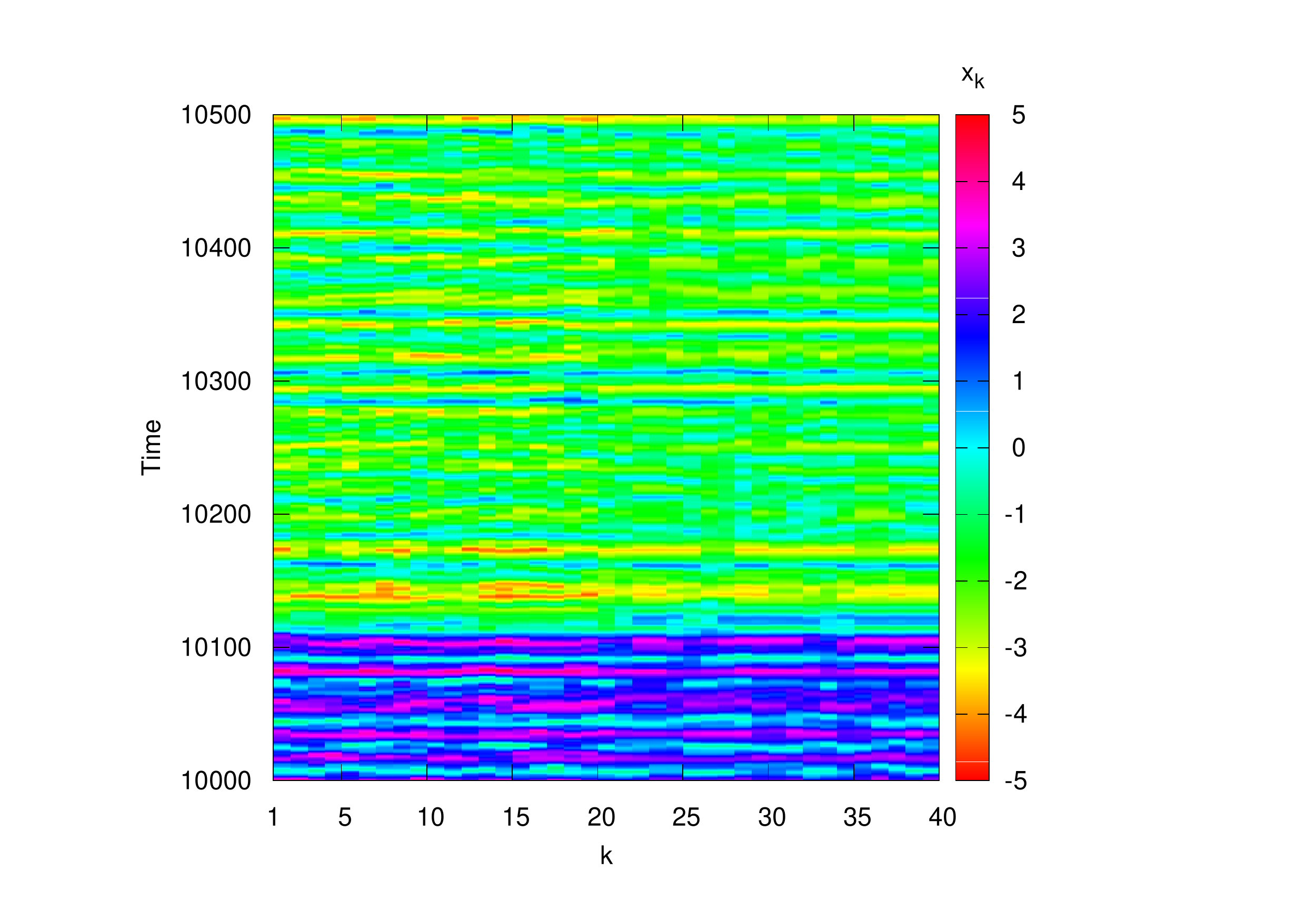}
\caption{}
\label{fig:array-vs-t_b00b018}
\end{subfigure}
\caption{Spatio-temporal phenomena of coherent and incoherent patterns, reminiscent of chimera states in 40 3-dimensional TR linearly coupled systems that exhibit labyrinth chaos and hyperchaos with $b_k=0$ for $k=1,\ldots,20$ (labyrinth chaos) and $b_k=0.18$ for $k=21,\ldots,40$ (hyperchaos).}
\label{fig:b00b018}
\end{figure}

Here, we focus on the case where the system in Eq. \eqref{eq:3DNPThomRoss} exhibits labyrinth chaos for half of the 3-dimensional TR systems and hyperchaos for the rest half. In particular, we set $N=40$ and $P=20$, $d=0.6$, $b_k=0$ for $k=1,\ldots,20$ (labyrinth chaos) and $b_k=0.18$ for $k=21,\ldots,40$ (hyperchaos) following Ren\'e and co-workers \cite{ThomRoss2004}. To identify the coherent and incoherent patterns of activity in the system, we compute at each time step of the numerical simulation, which $x_k$ values are locked
\begin{equation}
|x_i-x_j|<h
\end{equation}
for all $i,j=1,\ldots,40$, where $h=10^{-3}$ is a small threshold to detect locking. At each time step of the simulation, when locking is detected, the corresponding $x_k,\;k=1,\ldots,40$ values are recorded and the simulation proceeds to the next time step. When the simulation finishes, one gets a spatio-temporal phenomenon of coherent and incoherent patters such as those depicted in fig. \ref{fig:b00b018}(b). In this plot, we show all $x_k$ values in the time interval $[10^4,1.05\times10^4]$, where blue corresponds to relatively high $x_k$ value and, red, yellow and orange to relatively smaller $x_k$ values. It is evident there are coherent and incoherent groups of $x_k$ variables that alternate in time in a dynamical fashion. To show clearer these patterns, we plot 2 representative examples of spatio-temporal behaviour in panel (a) of the same figure where it is evident the existence of the coherent (locked) and incoherent groups of $x_k$ variables. The plots in panel (a) correspond to times $t=14462$ (upper plot) and $t=14515$ (lower plot). We have been able to observe similar patterns of spatio-temporal behaviour in other times as well.

\subsection{Chimera states in a multi-dimensional TR system with labyrinth chaos and complex periodic oscillations}

\begin{figure}[H]
\begin{subfigure}{0.45\textwidth}
\includegraphics[width=0.9\textwidth]{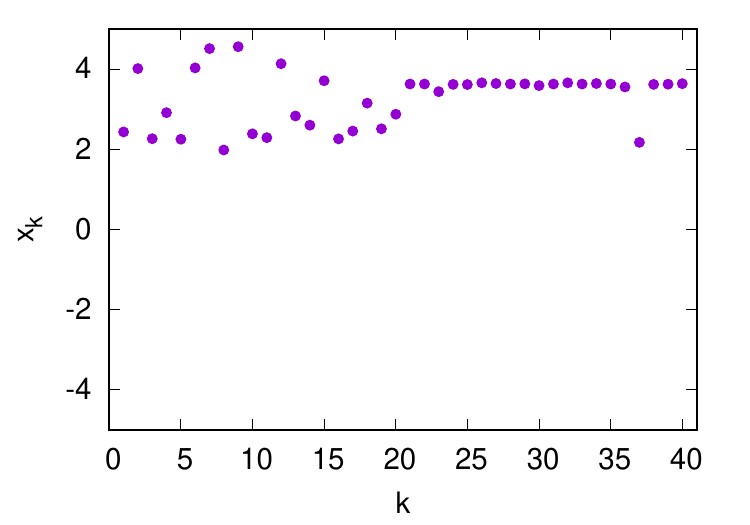}\\
\includegraphics[width=0.9\textwidth]{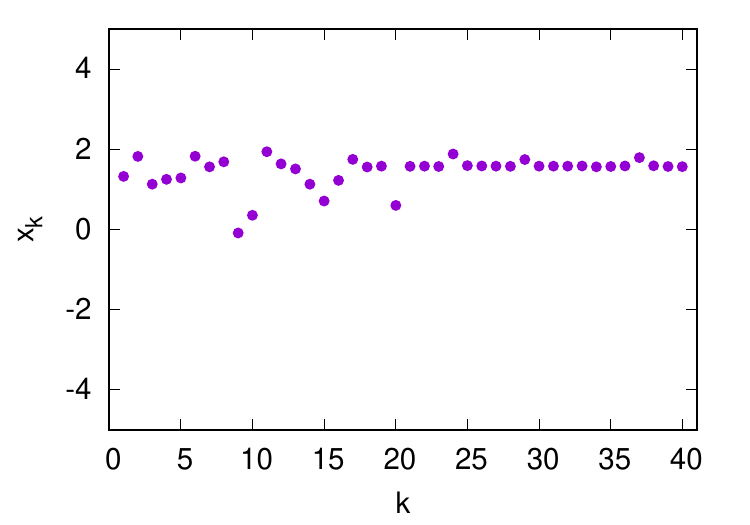}
\caption{}
\label{fig:profile_b00b019}
\end{subfigure}
\begin{subfigure}{0.5\textwidth}
\includegraphics[width=9.7cm,height=10cm,angle=-90]
{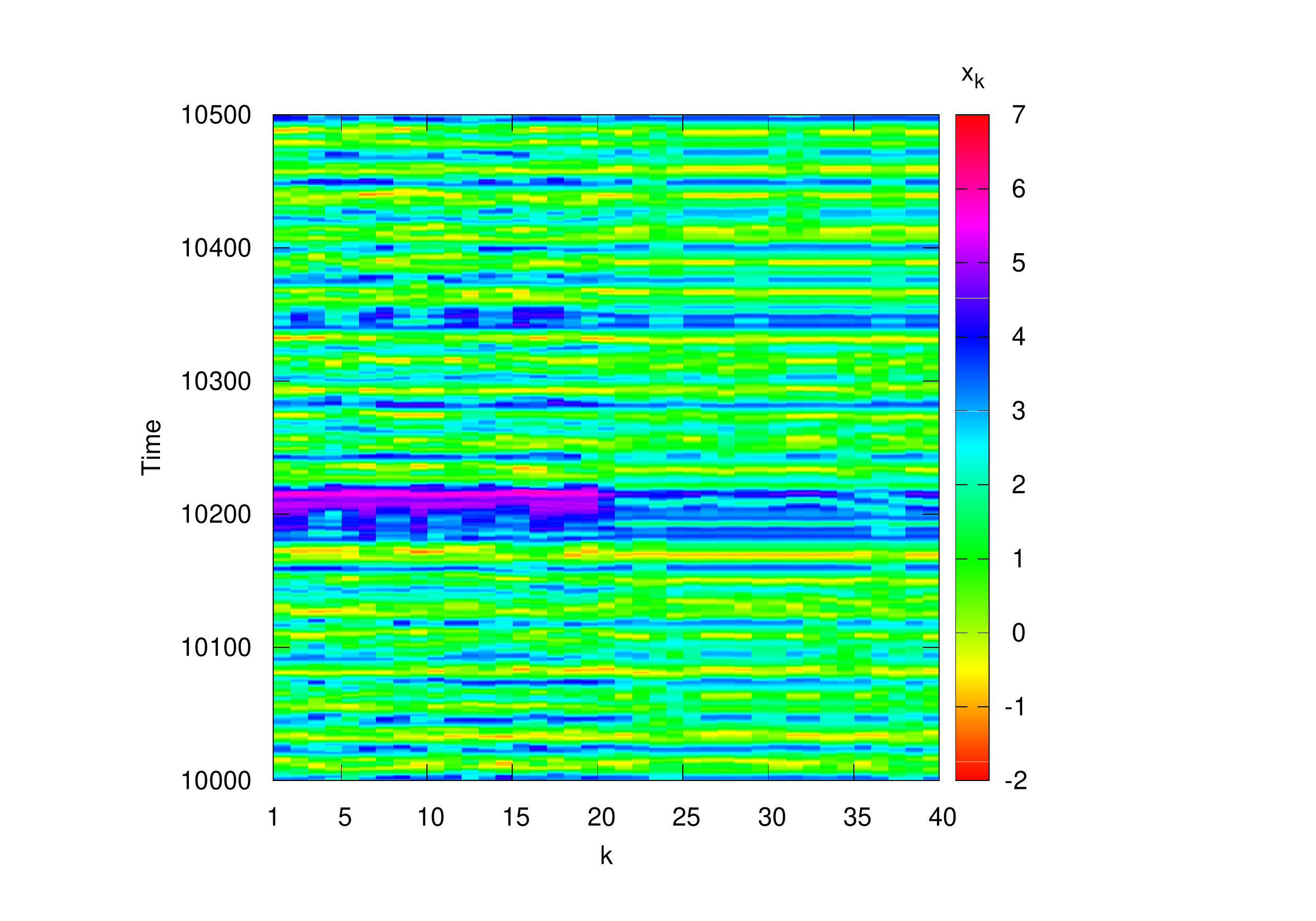}
\caption{}
\label{fig:array-vs-t_b00b019}
\end{subfigure}
\caption{Spatio-temporal phenomena of coherent and incoherent patterns, reminiscent of chimera states in 40 3-dimensional TR linearly coupled systems that exhibit labyrinth chaos and complex periodic oscillations with $b_k=0$ for $k=1,\ldots,20$ (labyrinth chaos) and $b_k=0.19$ for $k=21,\ldots,40$ (complex periodic oscillations).}
\label{fig:b00b019}
\end{figure}

Finally, we focus on the case where the system in Eq. \eqref{eq:3DNPThomRoss} exhibits labyrinth chaos for half of the 3-dimensional TR systems and complex periodic oscillations for the rest half. In particular, we set $N=40$ and $P=20$, $d=0.6$, $b_k=0$ for $k=1,\ldots,20$ (labyrinth chaos) and $b_k=0.19$ for $k=21,\ldots,40$ (hyperchaos) following again Ren\'e and co-workers \cite{ThomRoss2004}. We follow the same approach as previously to detect locking of the $x_k$ variables and plot in fig. \ref{fig:b00b019}(b) the spatio-temporal patterns of the activity of all $x_k$ values in the time interval $[10^4,1.05\times10^4]$. In this plot again blue corresponds to relatively high $x_k$ value and, red, yellow and orange to relatively smaller $x_k$. The coherent and incoherent patterns are again evident and alternate in time in a dynamical fashion. Figure \ref{fig:b00b019}(a) shows clearer these patterns where we plot 2 representative examples of spatio-temporal behaviour taken at 2 specific times from panel (b).  Again, there exists coherent (locked) and incoherent groups of $x_k$ variables that are reminiscent of chimera states. The plots in panel (a) correspond to times $t=10184$ (upper plot) and $t=10371$ (lower plot). As in the previous case, we have been able to observe similar patterns of spatio-temporal behaviour in other times as well.

 \section{Conclusions}\label{sect:conclusion}
During the last decade or so of his life, Ren\'e Thomas was joyfully preoccupied, among other things, with his Arabesque systems, his Labyrinth Chaos and his Chaotic Walkers. With this contribution to the volume, we had the opportunity to revisit his later work, where we had the honour to contribute to. We extended the TR systems to a spatio-temporal setting and we observed that it can support a behaviour which is reminiscent of chimera states for a wide range of parameter values of the linear coupling term.

We have here quite a novel case: the locking and drifting schemes are now due to labyrinth chaos and (hyper-)chaotic walks and not due to strict periodicity as in other cases in literature. This strongly suggests that our preliminary results in this paper rightfully asks for further detailed investigations as the role of coupling and dissipation are important. Here, we started with the simplest case where the $x_k$ variables of the Thomas-R\"ossler systems are linearly non-locally coupled. This, in effect, sets apart the $b$ parameter, which controls the dissipation for the $x_k$ variable, and changes it in time with the $(d/2P)x_k$ part of the coupling. This situation is indeed reminiscent of the case where Arabesques have different  dissipation parameters (like the $b$'s) for each variable. 

It seems that this is one of the last scientific avenues for further research Ren\'e opened, and it will lead to fruitful investigations and spectacular new results that will help shed light on the fundamentals of the emergence of chimera states. This will further provide importance on the mathematical modelling of biological significance. Ren\'e's work is so hopeful and inspirational, we are sure will lead to new fruitful investigations to elucidate the basic logic underlying chimera states, the outermost of complex dynamics.

\section*{Acknowledgements}
We dedicate this paper to the memory of our friend, Ren\'e Thomas, with our fondest memories. We wish to thank his good friends and our teachers O. E. R\"ossler and G. Nicolis. Also, M. Kaufman for her kind patience and for discussing in depth the logic of the ``feedback-approach'' all these years.


\section*{References}

\begin{thebibliography}{99}
% 
% \bibitem{Useless}
%  The Usefulness of Useless Knowledge
% by Abraham Flexner,
% Robbert Dijkgraaf (Commentary)
% 

\bibitem{ThomRoss2004}
R. Thomas, V. Basios, M. Eiswirth, T. Kruel and O. E. R\"ossler,
``Hyperchaos of arbitrary order generated by a single feedback circuit, and the emergence of 
chaotic walks'',  Chaos, 14, 3, 669-674, (2004).

\bibitem{ThomRoss2006}
R. Thomas and O. E. R\"ossler,
``Gen\`ese de Formes'', Revue des Questions Scientifiques, 177, 271-287, (2006). 

\bibitem{Arabesque2013}
C. Antonopoulos, V. Basios, J. Demongeot, P. Nardone and R. Thomas,
``Linear and nonlinear arabesques: a study of closed chains of negative 2-element circuits'',
Int. J. Bifurcation and Chaos, 23, (2013). 

\bibitem{Kaufman2013}
J. P. Comet, M. Noual, A. Richard, J. Aracena, L. Calzone, J. Demongeot, M. Kaufman, A. Naldi, E. H. 
Snoussi and D. Thieffry,
``On circuit functionality in Boolean networks'', Bull. Math. Biol. 75, 906-919, (2013).

\bibitem{Kaufman1995}
R. Thomas, D. Thieffry and M. Kaufman,
``Dynamical behaviour of biological regulatory networks.  I. 
Biological role of feedback loops and practical use of the concept of the loop-characteristic 
state'', Bull. Math. Biol., 57, 247-276, (1995).

\bibitem{Kaufman2001A}
R. Thomas and M. Kaufman,
``Multistationarity, the basis of cell differentiation and memory. I. Structural conditions of 
multistationarity and other non-trivial behaviour'', Chaos, 11, 
170-179, (2001).

\bibitem{Kaufman2001B}
R. Thomas and M. Kaufman,
``Multistationarity, the basis of cell differentiation and memory. II. Logical analysis of 
regulatory networks in terms of feedback circuits'', Chaos, 11, 180-195, (2001).

\bibitem{Kaufman2003complex}
M. Kaufman and R. Thomas,
``Emergence of complex behaviour from simple circuit structures'', 
C. R. Acad. Sci. Paris Biologies, 326, 205-214, (2003).

\bibitem{Scholarpedia}
C. Letellier and O. E. R\"ossler,
``Hyperchaos'', Scholarpedia, 2, 8, 1936, (2007).

\bibitem{Hyperchaos79}
O. E. R\"ossler,
``An equation for hyperchaos'', Physics Letters A, 71, 155-157, (1979).

\bibitem{Hyperchaos88}
F. T. Arecchi,
``Instabilities and chaos in lasers: Introduction to hyperchaos'',
In Order and Chaos in Nonlinear Physical Systems. Physics of Solids and Liquids, Springer, 
Boston, MA, (1988).

\bibitem{HyperchaosNavier}
R. A. Miranda, E. L. Rempel, A. C. Chian, N. Seehafer, B. A. Toledo and P. R. Mu\~noz,
``Lagrangian coherent structures at the onset of hyperchaos in the two-dimensional Navier-Stokes 
equations'',
Chaos, 23, 033107, (2013).

\bibitem{Hyperchaos17}
A. Arellano-Delgado, R. M. Lopez-Gutiérrez, M. A. Murillo-Escobar, L. Cardoza-Avenda\~no and C. Cruz-Hern\'andez,
``The emergence of hyperchaos and synchronisation in networks with discrete 
periodic oscillators'', Entropy, 19, 413, (2017).

\bibitem{HyperchaosDesign}
D, Silk, P. D.W. Kirk, C. P. Barnes, T. Toni, A. Rose, S. Moon, M. J. Dallman and M. P. H. Stumpf,
``Designing attractive models via automated identification of chaotic and oscillatory dynamical 
regimes.'', Nature Communications, 2, 489, (2011).

\bibitem{HyperchaosGenes}
V. A. Likhoshvai, S. I. Fadeev, V. Kogai and T. M. Khlebodarova,
``On The chaos in gene networks'',
Journal of Bioinformatics and Computational Biology, 11, 1, 1340009, (2013).

\bibitem{SprottChlouverakis}
J. C. Sprott and K. E. Chlouverakis,
``Labyrinth chaos'',
Int. J. Bifurcation and Chaos, 17,  6, 2097-2108, (2007).

\bibitem{SprottBook}
J. C. Sprott,
``Elegant chaos: Algebraically Simple Chaotic Flows'',
World Scientific, (2010).

\bibitem{LiSprott2014}
C. Li and J. C. Sprott,
``Coexisting hidden attractors in a 4-D simplified Lorenz system'', 
Int. J. Bifurcation Chaos; 24, 1450034, (2014).

\bibitem{BountisIranians2018}
F. Parastesh, S. Jafari, H. Azarnoush, B. Hatef and A. Bountis,
``Imperfect chimeras in a ring of 4-dimensional simplified Lorenz systems'',
Chaos, Solitons and Fractals, 110,  203-208, (2018).

\bibitem{HyperchaosChrono}
J. A. Betancourt-Mar, V. A. M\'endez-Guerrero, C. Hern\'andez-Rodr\'iguez and J. M. Nieto-Villar,
``Theoretical models for chronotherapy: Periodic perturbations in hyperchaos'',
Mathematical Biosciences and Engineering, 7, 3, 553-560, (2010).

\bibitem{HyperchaosNeurons}
P. Varona, M. I. Rabinovich, A.I . Selverston, and Y. I. Arshavsky,
``Winnerless competition between sensory neurons generates chaos: A possible
mechanism for molluscan hunting behaviour'',
Chaos, 12, 672, (2002).

\bibitem{LEs1980}
G. Benettin, L. Galgani, A. Giorgilli and J-M. Strelcyn,
``Lyapunov characteristic exponents for smooth dynamical systems and for 
Hamiltonian systems; A method for computing all of them''. Part 1: Theory; and Part 2: Numerical application,
Meccanica, 9 March and 21 March (1980).

\bibitem{kuramotoetal2002}
Y. Kuramoto and D. Battogtokh,
``Coexistence of coherence and incoherence in nonlocally coupled phase oscillators'',
Nonlinear Phenomena in Complex systems: An Interdisciplinary Journal, 5, 4, 380-385, (2002).

\bibitem{chimera-Imperfect}
T. Kapitaniak, P. Kuzma, J. Wojewoda, K. Czolczynski and Y. Maistrenko,
``Imperfect chimera states for coupled pendula'',
Scientific Reports, 4, 6379, (2014).

\bibitem{chimera-neurons}
A. Schmidt, T. Kasimatis, J. Hizanidis, A. Provata and P. H\"ove,
``Chimera patterns in two-dimensional networks of coupled neurons'',
Phys. Rev. E, 95, 032224, (2017).

\bibitem{chimera_Abrams}
D. M. Abrams and S. H. Strogatz,
``Chimera states for coupled oscillators'',
Phys. Rev. Lett., 93, 174102, (2004).

\bibitem{Panaggioetal2015}
M. J. Panaggio and D. M. Abrams,
``Chimera states: coexistence of coherence and incoherence in networks of coupled oscillators'',
Nonlinearity, 28, 3, (2015).

\end{thebibliography}
\end{document}